\documentclass[letterpaper]{article}

\usepackage{natbib,alifeconf}
\usepackage{graphicx}  
\usepackage{amsmath}
\usepackage{cleveref} 

\usepackage{breakcites}
\usepackage{array, caption, floatrow, tabularx, makecell, booktabs}%
\newfloatcommand{capbtabbox}{table}[][\FBwidth]
\setcellgapes{3pt}
\usepackage{subfiles}
\usepackage{times}
\usepackage{xcolor}
\usepackage{soul}
\usepackage[super]{nth}
\usepackage{geometry}
\usepackage[utf8]{inputenc}
\usepackage{textcomp}
\usepackage[utf8]{inputenc}
\usepackage{siunitx,booktabs}
\usepackage{times}  
\usepackage{helvet}  
\usepackage{courier}  
\usepackage{url}  
\usepackage{amssymb}

\graphicspath{ {eps/} }
\usepackage{booktabs}
\usepackage{latexsym} 
\usepackage{subcaption,siunitx,booktabs}
\usepackage{tikz}

\usepackage{caption}
\title{Modelling Cooperation in a Dynamic Healthcare System}
\author{Zainab Alalawi, The Anh Han, Yifeng Zeng \and  Aiman Elragig \\
\mbox{}\\
School of Computing and Digital Technologies, Teesside University, United Kingdom, TS1 3BX\\
z.alalawi@tees.ac.uk, t.han@tees.ac.uk, y.zeng@tees.ac.uk, a.elragig@tees.ac.uk} 

\begin{document}
\maketitle
The National Health Service (NHS) is a complex system that provides a free healthcare service at the point of delivery in the UK funded through taxpayers\textquotesingle~contributions \citep{slawson2018NHS}. Some of the NHS providers' budget is spent on sourcing services from Independent Sector Provider (ISP)/private sector.
Motivated by the NHS's {\it Five Year Forward View} (FYFV), we choose the patient as the core focus of healthcare planning who is to be included in the decision-making process for better health, patient care and financial sustainability. Most systematic studies looking into the improvement of clinical decision-making dealt with patients\textquotesingle~expectations of required treatments and their situations, and how that influences their clinical decision-making. However, considerations were slightly limited when it comes to understanding the patient\textquotesingle s role quantitatively as part of dynamic system modelling. 

Evolutionary Game Theory (EGT) have been applied to study various behavioral interactions between individuals and within dynamic systems across a wide range of disciplines running the gamut from economics, politics and security, to ecology, mathematical biology, and computer science \citep{nowak:ED,adami2016evolutionary}. In \cite{alalawi2019Path}, EGT is used to investigate the interaction between three populations, namely, the Public sector (P1) which represents the NHS or the public healthcare, the Private sector (P2) sells healthcare services, and the Patient (P3) represents a person seeking treatment(s). Several mechanisms are available to be used with the EGT; punishment is one mechanism that can enhance cooperation between individuals caught in social dilemmas \citep{hauert2007via}.
In this extended abstract, we outline how the basic model developed in our paper \citep{alalawi2019Path} is used to explain the interactions between the three populations.
The proposed model closely captures the costs and benefits of every strategy combining decisions by agents in a finite population on either {\it cooperation} or {\it defection}. We further introduce peer punishment into the model: the patients\textquotesingle~punishment which takes the form of complaints for clinical negligence \citep{bryden2011duty, cooper:2010}.

 Every individual or agent in each of the three populations experiences one of the following scenarios based on two strategies, namely, \textit{cooperate (C)} and \textit{defect (D)}. An individual from each population can choose one of the two strategies: provide/accept sustainable treatment(s) identified as cooperating, otherwise can't provide/refuse treatment(s) leading the patient to seek alternative treatment(s) from other providers. The payoff of each agent is acquired based on the selected strategy played by each agent from the three populations.
 
In \cite{alalawi2019Path}, our models (With/without Punishment) for any selected strategy help to understand how cooperation evolves in altruistic interactions among individuals of the interacting populations. Here, we assumed that the populations are of a fixed size \textit{N}. Individuals have the choice to either \textit{(C)} or \textit{(D)} in a paradigm shift fashion. 
In our developed model \citep{alalawi2019Path}, there are eight possible scenarios coinciding to the eight possible combinations of the basic strategies within the three populations, namely, \textit{CCC, CCD, CDC, CDD, DCC, DCD, DDC} and {\it DDD}.
The fractions of cooperators in P1, P2 and P3 is denoted by $x$, $y$, and $z$, respectively. The payoff of each strategy can be  written  as follows:~$P_s^{Population}(x,y,z) = P_{syz}$, where $x,y,z\in \{0,1\}$.
For instance, individuals from the population  P1 have the option to play \textit{C} or \textit{D}. The selected strategy will replace the \textit{s} at \textit{x} vertex, while \textit{y} and \textit{z} vertices remain unchanged for every selected strategy for the Public population.
The payoff of randomly selected individuals \textit{A} and \textit{B} in the population depends on the proportion of both players in the population.

We assume that at any time-step, an individual \textit{B} with fitness $\pi_B$ imitates a randomly selected individual \textit{A} with $\pi_A$ fitness adopting a pairwise comparison rule \citep{nowak:ED}. The  probability $\rho$ of A adopts B\textquotesingle s strategy is defined by Fermi\textquotesingle s function $\rho = {[1+ e^{-\beta[\pi_A - \pi_B]}]}^{-1}$ \citep{sigmund2010social, traulsen2007pairwise}. Here, $\beta$ represents the \lq intensity of selection\rq ($\beta = 0$ represents neutral drift and close to natural selection when it is small, while for large $\beta \rightarrow \infty$ the imitation decision is increasingly deterministic).
{\small
\begin{figure}[!t]
\includegraphics[width=\textwidth]{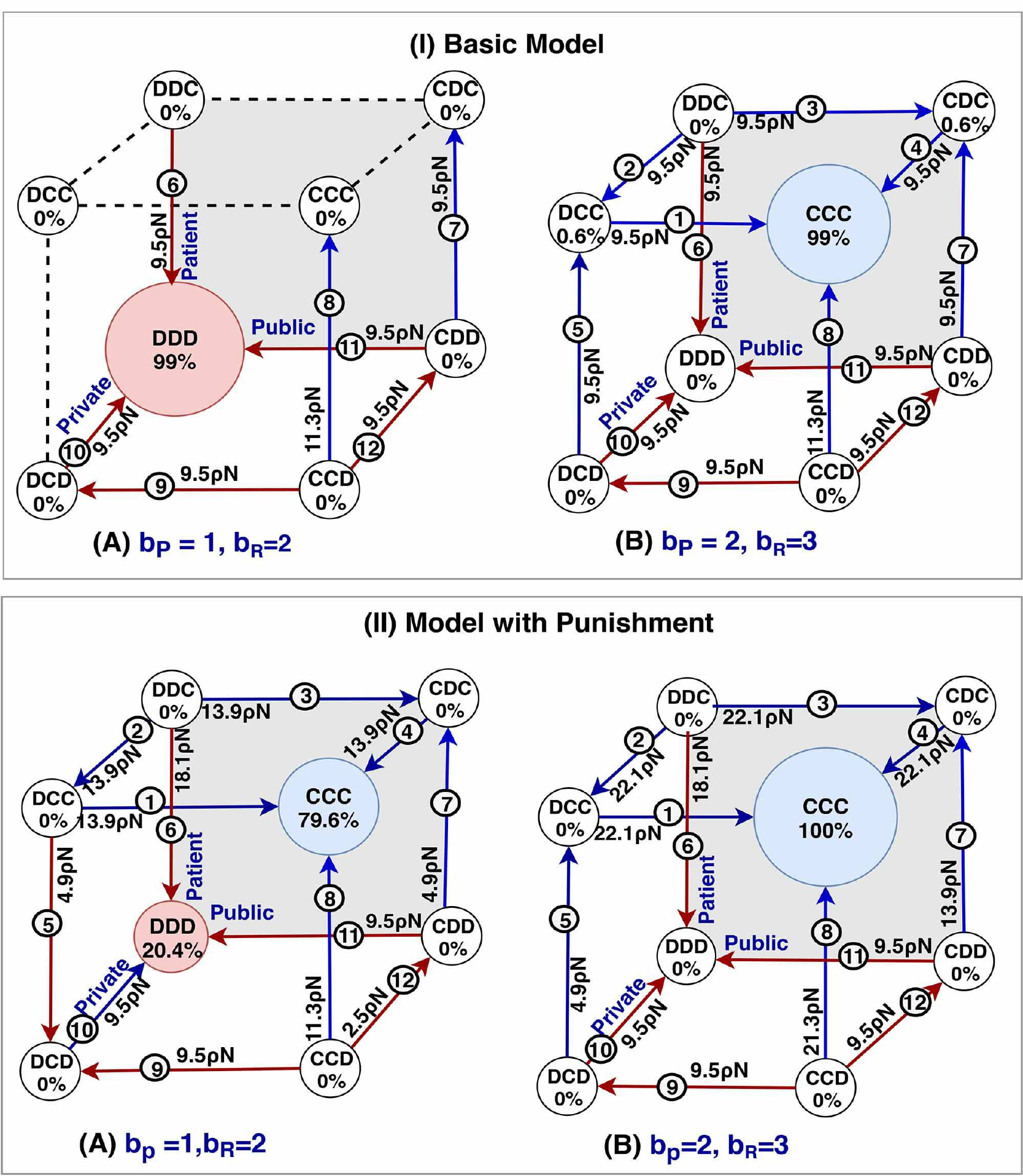}
\caption{Stationary distribution and fixation probabilities. It illustrates the parameters values as stated for each simplex, whereas: red arrow represents transition towards defection, blue arrow transition towards cooperation and dashed line refers to neutral state. A high percentages 80\% of individuals would adopt cooperation with lower $b_P$ in (II-A) compared to (I-A), whilst an adequate 1\% increase in cooperation with higher $b_R$ in (II-B) compared to (I-B) is registered. The transition probability and frequency dependency are normalised ($1/N$), where $N=100$. Other parameters: $c_I, c_T, c_M=1, \varepsilon=0.2$ and $\beta=0.1$.}
\label{fig:SDResult}
\end{figure}}
A symmetric matrix is constructed by computing the average fitness of an individual adopting a strategy \textit{s} within a population that is obtained from the tripartite one-shot game described in Table~1 see \citep{alalawi2019Path}. 

A numerical and systematic analysis is performed on the basic model, we focus on the social interactions between players in each population and how their decision influences the level of cooperation to sustain cost-effective services and better patient satisfaction. By simulating the models with a selected range of parameters ($b_P$ and $b_R$), (Fig. \ref{fig:SDResult}I-A) shows that the \textit{DDD} strategy dominates the populations' dynamic by 99\%. As it has been observed in the basic model where punishment is absent, players of each population spend most of their time at the state of defecting strategies (see Fig.~\ref{fig:SDResult} I-A). We started by pairwise computation of the interaction strategies in the payoff matrix (Table~2) in \cite{alalawi2019Path} based on different values of the parameters ($b_R$ and $b_P$) to measure the stationary distribution and the frequency of the eight strategies.
\begin{figure}[!hbt]
\includegraphics[width=\textwidth]{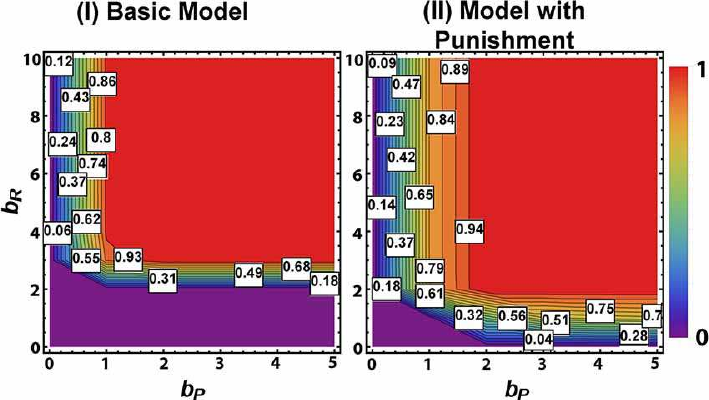}
\caption{Frequency of strategy \textit{CCC} for varying the main parameters $b_R$ and $b_P$:~(I) for the basic model, (II) extended model with costly punishment. In (II) $u =0.5$ and $v=1.5$. In general, \textit{CCC} performs better when punishment is introduced to the basic model. Additionally, significant cooperation is achieved for sufficiently low $b_R$ and large $b_P$. Other parameters $c_I,c_T,c_M =1, \varepsilon=0.2, N=100,$ and $\beta=0.1$.}
\label{fig:ControuPlot1}
\end{figure}
Based on our proposed model, the patient has the option to mete out a costly punishment \textit{v} to the defecting healthcare provider(s) at~\textit{u} cost (i.e. legal fees).
The most dominant strategy (i.e, \textit{CCC} - full cooperation) is where the most finite populations reside. When the patient\textquotesingle s punishment is applied (see Fig.~\ref{fig:SDResult}~II-A~$\&$~B), cooperators invade defectors when the acquired patient\textquotesingle s benefit (namely, $b_P > 1$) is large enough, and the populations spend at least 79.6\% of their time in cooperation. Our simulation suggests that, despite the low reputation\textquotesingle s benefit, cooperation is highly achievable (see Fig. \ref{fig:SDResult} II). The results of predicted frequencies for both models illustrated in (Fig. \ref{fig:ControuPlot1} I$\&$II) show a significant increase in cooperation where ($b_R > 1$ and $b_P>1$).

Future work will involve investigating the implementation of institutional punishment. Furthermore, we will introduce new factors (e.g., waiting time) to investigate the if there are any other drivers leading to cooperation. Finally, our proposed model directs attention to how the patient's decision would impact the process of collaboration between healthcare providers, and to the effectiveness of management decisions made by the private sector in influencing the patient\textquotesingle s choice of cooperation.

\bibliographystyle{apalike}

\fontsize{8pt}{9pt}\selectfont
\bibliography{main} 

\end{document}